\documentstyle[11pt]{article}
\def\wh{\hat{w}}
\def\beq{\begin{equation}}
\def\eeq{\end{equation}}
\def\wn{w_{_{\rm N}}}
\def\we{w_{_{\rm E}}}
\def\ws{w_{_{\rm S}}}
\def\ww{w_{_{\rm W}}}
\def\un{u_{_{\rm N}}}
\def\ue{u_{_{\rm E}}}
\def\us{u_{_{\rm S}}}
\def\uw{u_{_{\rm W}}}
\def\wtn{\tilde{w}_{_{\rm N}}}
\def\wte{\tilde{w}_{_{\rm E}}}
\def\wts{\tilde{w}_{_{\rm S}}}
\def\wtw{\tilde{w}_{_{\rm W}}}
\def\psin{\psi_{_{\rm N}}}
\def\psie{\psi_{_{\rm E}}}
\def\psis{\psi_{_{\rm S}}}
\def\psiw{\psi_{_{\rm W}}}
\def\psia{\psi_{_{\rm A}}}
\def\psib{\psi_{_{\rm B}}}
\def\chin{\chi_{_{\rm N}}}
\def\chie{\chi_{_{\rm E}}}
\def\chis{\chi_{_{\rm S}}}
\def\chiw{\chi_{_{\rm W}}}
\def\varphin{\varphi_{_{\rm N}}}
\def\varphie{\varphi_{_{\rm E}}}
\def\varphis{\varphi_{_{\rm S}}}
\def\varphiw{\varphi_{_{\rm W}}}
\def\U{{\bf U}}
\def\L{{\bf L}}
\def\l{\lambda}
\def\half{\frac{1}{2}}
\def\tr{{\rm tr}}

\begin{document}

\bigskip
\begin{center}
{\Large\bf
Hirota equation as an example of integrable
symplectic map}\footnote[1]
{To appear in Lett. Math. Phys.}\\
\bigskip
{\large\bf
     L. Faddeev$^{1,}$\footnote[2]
     {Supported by the
 Russian Academy of Sciences and the Academy of Finland}
     and A. Yu. Volkov$^{2,}$\footnote[3]
     {On leave of absence from
Saint Petersburg Branch of the Steklov Mathematical Institute,\\
Fontanka 27, Saint Petersburg 191011, Russia}}
\end{center}
\bigskip
\begin{quote}
$^1$ {\em Saint Petersburg Branch of the Steklov Mathematical Institute\\
     Fontanka 27, Saint Petersburg 191011, Russia\\{\em and}
     Research Institute for Theoretical Physics\\
 P. O. Box 9 (Siltavuorenpenger 20C), SF-00014
 University of Helsinki, Finland\\
$^2$ Physique - Math\'{e}matique, Universit\'{e} de Bourgogne\\
     B.P.138, 21004 Dijon C\'{e}dex, France}

\bigskip
{\bf Abstract.}The hamiltonian formalism is developed for the Sine-Gordon
 model on the space-time light-like lattice, first introduced by Hirota.
 The evolution operator is explicitely constructed in the quantum variant
 of the model, the integrability of corresponding classical
 finite-dimensional system is established.
\end{quote}

\section*{Introduction}

We consider a system of difference equations on a plane lattice
shown below.
$$
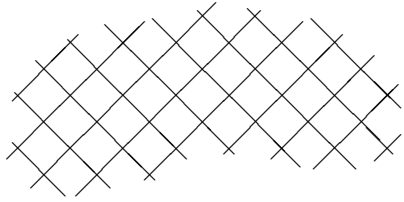
\begin{picture}(160,80)
\put(8,38){\line(1,1){25}}
\put(6,16){\line(1,1){52}}
\put(15,5){\line(1,1){70}}
\put(32,2){\line(1,1){70}}
\put(58,8){\line(1,1){60}}
\put(88,18){\line(1,1){45}}
\put(106,16){\line(1,1){39}}
\put(122,12){\line(1,1){32}}
\put(145,15){\line(1,1){10}}
\put(8,22){\line(1,-1){20}}
\put(9,41){\line(1,-1){36}}
\put(17,53){\line(1,-1){45}}
\put(29,61){\line(1,-1){45}}
\put(43,67){\line(1,-1){49}}
\put(61,69){\line(1,-1){56}}
\put(78,72){\line(1,-1){60}}
\put(97,73){\line(1,-1){55}}
\put(121,69){\line(1,-1){34}}
\end{picture}$$
The variables, attached to the vertices of lattice, are denoted
by $w$. The equations connect the values
of $w$'s belonging to a common elementary square
$$
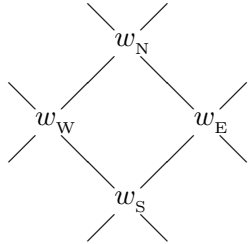
\begin{picture}(100,100)
\put(5,35){\line(1,1){10}}
\put(25,55){\line(1,1){20}}
\put(55,85){\line(1,1){10}}
\put(35,5){\line(1,1){10}}
\put(55,25){\line(1,1){20}}
\put(85,55){\line(1,1){10}}
\put(5,65){\line(1,-1){10}}
\put(25,45){\line(1,-1){20}}
\put(55,15){\line(1,-1){10}}
\put(35,95){\line(1,-1){10}}
\put(55,75){\line(1,-1){20}}
\put(85,45){\line(1,-1){10}}
\put(15,49){$\ww$}
\put(45,19){$\ws$}
\put(45,79){$\wn$}
\put(75,49){$\we$}
\end{picture}$$
as follows
\beq
   \wn=\ws\frac{f(\ww)}{f(\we)}.
\label{eq:hirota}
\eeq
The equation
allows to get $\wn$ if three other $w$'s are given and so defines
an evolution in the vertical direction (from south to north).

If the function $f(w)$ is invertible, so that the equation
\beq
     \tilde{w}=f(w)
\eeq
has a unique solution
\beq
     w=g(\tilde{w})
\eeq
then the equation (\ref{eq:hirota}) can be rewritten in the form
\beq
     \wtw=\wte\frac{g(\wts)}{g(\wtn)}
\eeq
and defines the evolution from east to west. In what follows we shall
consider only move south-north, defining the time axis as vertical.
Then the horisontal saw
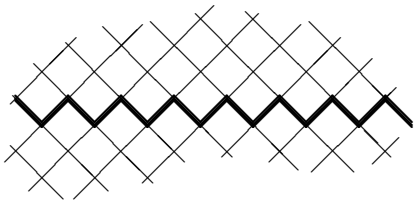
$$\begin{picture}(160,80)
\put(8,38){\line(1,1){25}}
\put(6,16){\line(1,1){52}}
\put(15,5){\line(1,1){70}}
\put(32,2){\line(1,1){70}}
\put(58,8){\line(1,1){60}}
\put(88,18){\line(1,1){45}}
\put(106,16){\line(1,1){39}}
\put(122,12){\line(1,1){32}}
\put(145,15){\line(1,1){10}}
\put(8,22){\line(1,-1){20}}
\put(9,41){\line(1,-1){36}}
\put(17,53){\line(1,-1){45}}
\put(29,61){\line(1,-1){45}}
\put(43,67){\line(1,-1){49}}
\put(61,69){\line(1,-1){56}}
\put(78,72){\line(1,-1){60}}
\put(97,73){\line(1,-1){55}}
\put(121,69){\line(1,-1){34}}
\thicklines
\multiput(20,30)(20,0){7}{\line(1,1){10}}
\multiput(10,40)(20,0){8}{\line(1,-1){10}}
\multiput(20,29)(20,0){7}{\line(1,1){10}}
\multiput(10,39)(20,0){8}{\line(1,-1){10}}
\multiput(20,31)(20,0){7}{\line(1,1){10}}
\multiput(10,41)(20,0){8}{\line(1,-1){10}}
\thinlines
\end{picture}$$
can be considered as a ``line" of initial data.

It is evident that the initial data, given on a portion of that
saw, define a solution on the part of a lattice bounded by the
punctured lines:
$$
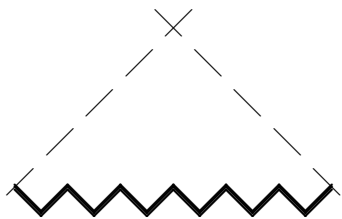
\begin{picture}(130,90)
\multiput(2,12)(15,15){5}{\line(1,1){10}}
\multiput(58,82)(15,-15){5}{\line(1,-1){10}}
\thicklines
\multiput(15,5)(20,0){6}{\line(1,1){10}}
\multiput(5,15)(20,0){6}{\line(1,-1){10}}
\multiput(15,4)(20,0){6}{\line(1,1){10}}
\multiput(5,14)(20,0){6}{\line(1,-1){10}}
\multiput(15,6)(20,0){6}{\line(1,1){10}}
\multiput(5,16)(20,0){6}{\line(1,-1){10}}
\thinlines
\end{picture}$$
In this sense we can speak of hyperbolicity of our system, the lines
being lightlike and defining the light cone.

Alternatively we can use a kind of dual variables $\psi$ to define
the same dynamical system. These variables are attached to the
vertices of a ``dual" lattice
$$
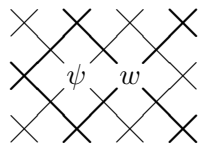
\begin{picture}(80,60)
\put(5,5){\line(1,1){20}}
\put(35,35){\line(1,1){20}}
\put(5,45){\line(1,1){10}}
\put(45,5){\line(1,1){30}}
\put(5,55){\line(1,-1){20}}
\put(35,25){\line(1,-1){20}}
\put(5,15){\line(1,-1){10}}
\put(45,55){\line(1,-1){30}}
\thicklines
\put(25,5){\line(1,1){20}}
\put(55,35){\line(1,1){20}}
\put(65,5){\line(1,1){10}}
\put(5,25){\line(1,1){30}}
\put(25,55){\line(1,-1){20}}
\put(55,25){\line(1,-1){20}}
\put(65,55){\line(1,-1){10}}
\put(5,35){\line(1,-1){30}}
\thinlines
\put(46,27){$w$}
\put(26,27){$\psi$}
\end{picture}$$
The connection of $\psi$'s and $w$'s is given by
$$\begin{picture}(80,80)
\thicklines
\put(5,5){\line(1,1){25}}
\put(50,50){\line(1,1){25}}
\put(5,65){\line(1,1){10}}
\put(65,5){\line(1,1){10}}
\put(5,75){\line(1,-1){25}}
\put(50,30){\line(1,-1){25}}
\put(5,15){\line(1,-1){10}}
\put(65,75){\line(1,-1){10}}
\thinlines
\put(5,39){$\psiw$}
\put(35,9){$\psis$}
\put(35,69){$\psin$}
\put(65,39){$\psie$}
\put(30,38){$w,\tilde{w}$}
\end{picture}$$
\beq
           w=\frac{\psiw}{\psie}
\eeq
and
\beq
   \psi=\prod w
\eeq
where the product is taken over all $w$'s on the same horisontal line to
the east of $\psi$.
\pagebreak
The second formula needs clarification of the boundary
conditions; this will be done in the main text for the lattice periodic
in the horisontal (spacial) direction.

In terms of $\psi$'s the equations take the form
\beq
  \psin=\psis f(\frac{\psiw}{\psie})
\label{eq:psi}
\eeq
which is also eqivalent to
\beq
       \tilde{w}=\frac{\psin}{\psis}
\eeq
(see the last picture).

In Section 1 we present a symplectic structure on the space of initial
data which is conserved by evolution and construct the generator of
this evolution. We shall do the latter in a quantum version of our system
where this generator is given by an unitary operator. Then in Section 2
for a concrete choice of the function $f$
\beq
  f(w)=\frac{1+\kappa^2 w}{\kappa^2+w}
\eeq
where $\kappa^2$ is a real positive parameter we shall show that our
symplectic map is completely integrable. Namely, we shall present the set
of independent commuting integrals of motion, as many as the number
of degrees of freedom. Finally in Section 3 we shall show the equivalence
of our system to that introduced by Hirota \cite{H} as the one
approximating the famous sine-Gordon equation. Some general comments are
collected in Conclusion. Some results of the present paper appeared
already in \cite{VF}.

\section{Symplectic map}
We shall consider our system on a spacially periodic lattice of even
length $2M$. The variables $w$ on a saw of initial data will be numbered
as shown on the picture:
$$
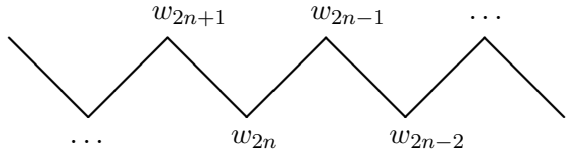
\begin{picture}(220,60)
\thicklines
\multiput(35,15)(60,0){3}{\line(1,1){30}}
\multiput(5,45)(60,0){4}{\line(1,-1){30}}
\thinlines
\put(28,5){$\ldots$}
\put(59,52){$w_{2n+1}$}
\put(89,5){$w_{2n}$}
\put(119,52){$w_{2n-1}$}
\put(149,5){$w_{2n-2}$}
\put(179,52){$\ldots$}
\end{picture}$$
We introduce the following Poisson structure
\beq \begin{array}{rcl}
    \{w_{n-1},w_n\} & = & 2\gamma w_{n-1}w_n \;\;\;\;
                         n=2,3,\ldots,2M  \\
    \{w_{2M},w_1\}  & = & 2\gamma w_{2M}w_1\\
    \{w_m,w_n\}     & = & 0 \;\;\;\;\;\;\;\;\;\; 1<|m-n|<2M-1.
     \end{array}
\label{eq:pb}
\eeq
Here $\gamma$ is a positive parameter playing the role of a coupling
constant. The Jacoby identity is evident as the Poisson matrix becomes
constant after passing to logarithms of $w$'s.

In applications one can encounter different variants for the set of
values of the variables $w$. The most evident are the following ones:\\
\indent 1) $w$ real and positive;\\
\indent 2) $w$ complex and unimodular, $|w|=1$.\\
Poisson brackets are compatible with both cases; there are also other
possibilities but in this paper we confine ourselves to the second
(compact) form of $w$. It is thus necessary to require that the function
$f$ maps the unit circle on itself.

The Poisson structure is degenerate; the functions
\beq \begin{array}{rcl}  C_1&=&\prod w_{odd}   \\
                         C_2&=&\prod w_{even}
     \end{array}
\label{eq:c}
\eeq
are commuting with all $w$'s. The submanifolds with fixed values of
$C_1$ and $C_2$ realize the symplectic leaves of our manifold. The
number of degrees of freedom is thus $M-1$.

The map
$$   {\cal U}:\;\; w\longrightarrow \hat{w}   $$
where $\hat{w}$'s are the values of the solution of the equations
(\ref{eq:hirota}) on a saw $\hat{S}$ shifted by time unit from
the initial saw $S$
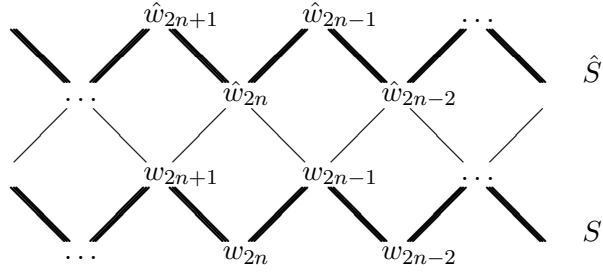
$$\begin{picture}(230,120)
\thicklines
\multiput(40,20)(60,0){3}{\line(1,1){20}}
\multiput(10,40)(60,0){4}{\line(1,-1){20}}
\multiput(40,80)(60,0){3}{\line(1,1){20}}
\multiput(10,100)(60,0){4}{\line(1,-1){20}}
\multiput(39,20)(60,0){3}{\line(1,1){20}}
\multiput(9,40)(60,0){4}{\line(1,-1){20}}
\multiput(39,80)(60,0){3}{\line(1,1){20}}
\multiput(9,100)(60,0){4}{\line(1,-1){20}}
\multiput(41,20)(60,0){3}{\line(1,1){20}}
\multiput(11,40)(60,0){4}{\line(1,-1){20}}
\multiput(41,80)(60,0){3}{\line(1,1){20}}
\multiput(11,100)(60,0){4}{\line(1,-1){20}}
\thinlines
\multiput(40,70)(60,0){3}{\line(1,-1){20}}
\multiput(10,50)(60,0){4}{\line(1,1){20}}
\put(29,13){$\ldots$}
\put(29,73){$\ldots$}
\put(59,43){$w_{2n+1}$}
\put(59,103){$\wh_{2n+1}$}
\put(89,13){$w_{2n}$}
\put(89,73){$\wh_{2n}$}
\put(119,43){$w_{2n-1}$}
\put(119,103){$\wh_{2n-1}$}
\put(149,13){$w_{2n-2}$}
\put(149,73){$\wh_{2n-2}$}
\put(179,43){$\ldots$}
\put(179,103){$\ldots$}
\put(225,20){$S$}
\put(225,80){$\hat{S}$}
\end{picture}$$
preserves the Poisson brackets and the values of invariants $C_1$ and
$C_2$. Thus ${\cal U}$ defines a symplectic map (canonical
transformation). It can be described by means of corresponding generating
function. We shall not follow this route here and instead shall describe
the unitary evolution operator $\U$ corresponding to this map in the
quantized variant of our dynamical problem. The reason for this is, that
quantum mechanics being more algebraic, the corresponding formulae
are just
more simple. One can then get the classical variant by the appropriate
limit.

After quantization the variables $w$ become unitary operators with the
commutation relations
\beq  \begin{array}{rcl}
    w_{n-1} w_n & = & q^2 w_n w_{n-1} \;\;\;\;
                         n=2,3,\ldots,2M  \\
    w_{2M} w_1  & = & q^2 w_1 w_{2M}       \\
    w_m w_n     & = & w_n w_m \;\;\;\;\;\;1<|m-n|<2M-1.
     \end{array}
\label{eq:w}
\eeq
where
$$  q={\rm e}^{i\hbar\gamma}   $$
$\hbar$ being the Planck constant. This commutation relation passes into
the Poisson relation (\ref{eq:pb}) with $\hbar\rightarrow 0$ if we
identify as usually
$\frac{i}{\hbar}[\;,\;]=\{\;,\;\}$.

The evolution operator is given by
\beq
    \U=\prod r(w_{even})\prod r(w_{odd})
\label{eq:u}
\eeq
where the function $r(w)$ is defined by the function $f(w)$ through the
functional equation
\beq
    \frac{r(qw)}{r(q^{-1}w)}=f(w).
\label{eq:feq}
\eeq
There is no ordering of factors problem here as only one variable
enters the functional equation while in (\ref{eq:u}) we took into
account that all even $w$'s commute among themselves, the
same for odd ones.

Application of $\U$ to $w$'s is rather trivial due to the locality
of their commutation relations. For even $w$'s we get
\beq
   \hat{w}_{2n}=\U^{-1} w_{2n} \U =
   r^{-1}(w_{2n-1})r^{-1}(w_{2n+1})w_{2n}r(w_{2n+1})r(w_{2n-1})
\eeq
which after using the functional equation (\ref{eq:feq}) gives
\beq
   \hat{w}_{2n}=f(qw_{2n+1})w_{2n}(f(qw_{2n-1}))^{-1}.
\eeq
For odd $w$'s the calculation consists of several steps but it is
as trivial as above. We get
\beq
     \hat{w}_{2n-1}=\U^{-1} w_{2n-1} \U =
     f(q\hat{w}_{2n})w_{2n-1}(f(q\hat{w}_{2n-2}))^{-1}.
\eeq
So the iteration of $\U$ leads to the quantum form of the equations of
motion
\beq
   \wn=f(q\ww)\ws(f(q\we))^{-1}
\eeq
which coincides with (\ref{eq:hirota}) in the classical
limit $q\rightarrow 1$.

Consider now the alternative picture in the variables $\psi$. The set
of variables $\psi_n$
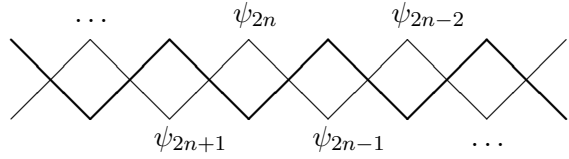
$$\begin{picture}(220,60)
\thicklines
\multiput(35,15)(60,0){3}{\line(1,1){30}}
\multiput(5,45)(60,0){4}{\line(1,-1){30}}
\thinlines
\multiput(5,15)(60,0){4}{\line(1,1){30}}
\multiput(35,45)(60,0){3}{\line(1,-1){30}}
\put(29,52){$\ldots$}
\put(59,5){$\psi_{2n+1}$}
\put(89,52){$\psi_{2n}$}
\put(119,5){$\psi_{2n-1}$}
\put(149,52){$\psi_{2n-2}$}
\put(179,5){$\ldots$}
\end{picture}$$
will be now quasiperiodic
\beq  \begin{array}{rcl}
            \psi_{2n+2M}&=&C_1 \psi_{2n}    \\
            \psi_{2n-1+2M}&=&C_2 \psi_{2n-1+2M}
      \end{array}
\eeq
and $2M+2$ independent variables $C_1,C_2,\psi_1,\psi_2,\ldots,\psi_{2M}$
($M+1$ degrees of freedom) have the following system of
commutation relations:
\beq \begin{array}{rcl}
           \psi_m\psi_n&=&q\psi_n\psi_m \;\;\;{\rm if}\;\;\;
                          n-m\;{\rm odd\;and}\;0<n-m<2M           \\
           \psi_{odd}C_1&=&q^2C_1\psi_{odd}                       \\
           \psi_{even}C_2&=&q^2C_2\psi_{even}  \\&&\\
    {[}\psi_{odd},\psi_{odd}{]}&=&{[}\psi_{even},\psi_{even}{]}\;=\;0.
     \end{array}
\eeq
Introduce the evolution by means of the same operator $\U$ with
\beq
      w_n=\frac{\psi_{n+1}}{\psi_{n-1}}.
\eeq
Of course these $w$'s define system (\ref{eq:w}) with $C_1$ and $C_2$ as
monodromies (\ref{eq:c}).

The derivation of equations of motion for $\psi$'s is even simpler
than that for $w$'s because each $\psi$ does not commute with only
one $w$
\beq
      \psi_n w_n=q^2 w_n \psi_n.
\eeq
We get
\beq
     \psin=\psis f(q^{-1}\frac{\psiw}{\psie})
\label{eq:qh}
\eeq
which passes to (\ref{eq:psi}) in the classical limit $q\rightarrow 1$.

\section{Integrability}

We consider now the particular case with the function $f(w)$ given by
\beq
  f(w)=\frac{1+\kappa^2 w}{\kappa^2+w}.
\eeq
Here $\kappa^2$ is a fixed positive parameter. We thus have $|f|=1$
if $|w|=1$. The inverse function $g$ is given by the same formula
after reflection $\kappa^2\rightarrow -\kappa^2$.

Of course this function has a very special character and it is not clear
{\em a priori} why it is interesting. Its exclusiveness is due to the
following property \cite{V}: the solution of the functional
equation
\beq
   \frac{r(\l,qw)}{r(\l,q^{-1}w)}=\frac{1+\l w}{\l+w}
\label{eq:fe}
\eeq
which we use in the definition of the evolution operator (\ref{eq:u})
\beq
    \U=\prod r(\kappa^2,w_{even})\prod r(\kappa^2,w_{odd})
\eeq
satisfies the Yang-Baxter equation in the following form: let $u$ and $v$
be a Weyl pair
\beq
   uv=q^2vu,
\eeq
then
\beq
   r(\lambda,u)r(\lambda\mu,v)r(\mu,u)=
    r(\mu,v)r(\lambda\mu,u)r(\lambda,v).
\eeq
This equation allows to interprete $r$ as a fundamental $L$-operator of
quantum chain \cite{TTF} and use the machinery of
QISM (see e.g. \cite{FLO}).
In particular we can introduce a family of matrix $L$-operators
with operator matrix elements
\beq
   L_{n-\half}(\l)=\left( \begin{array}{cc}
\psi_{n-1}^{-\half}\psi_{n}^{\half}&
\l\psi_{n-1}^{-\half}\psi_{n}^{-\half}\\
                               &\\
\l\psi_{n-1}^{\half}\psi_{n}^{\half}&\psi_{n-1}^{\half}\psi_{n}^{-\half}
   \end{array} \right).
\eeq
Due to (\ref{eq:fe}) it is connected with $r$ by the commutation
relation \cite{V}
\beq
  r({\textstyle \frac{\l}{\mu}},w_n)L_{n+\half}(\l)L_{n-\half}(\mu)=
  L_{n+\half}(\mu)L_{n-\half}(\l)r({\textstyle \frac{\l}{\mu}},w_n)
\label{eq:cr}
\eeq
where presumably $w_n=\frac{\psi_{n+1}}{\psi_{n-1}}$.

One can recognize in $L$ a Lax operator for the Hirota
equation. Indeed, the equation (\ref{eq:qh}) could be
interpreted as a zero curvature condition
for the elementary square of our lattice:
$$\begin{picture}(100,105)
\put(35,80){\line(1,1){10}}
\put(80,35){\line(1,1){10}}
\put(35,20){\line(1,-1){10}}
\put(80,65){\line(1,-1){10}}
\put(20,35){\vector(-1,1){10}}
\put(20,65){\vector(-1,-1){10}}
\put(65,80){\vector(-1,1){10}}
\put(65,20){\vector(-1,-1){10}}
\put(0,49){$\psiw$}
\put(90,49){$\psie$}
\put(45,4){$\psis$}
\put(45,94){$\psin$}
\put(-10,69)
{$L_{\rm W\leftarrow N}(\kappa^{-1}\l)\;\;\;\;
 L_{\rm N\leftarrow E}(\kappa\l)$}
\put(-10,24)
{$L_{\rm W\leftarrow S}(\kappa\l)\;\;\;\;\;\,
 L_{\rm S\leftarrow E}(\kappa^{-1}\l)$}
\end{picture}$$
\beq
L_{\rm W\leftarrow N}(\kappa^{-1}\l)L_{\rm N\leftarrow E}(\kappa\l)=
L_{\rm W\leftarrow S}(\kappa\l)L_{\rm S\leftarrow E}(\kappa^{-1}\l)
\label{eq:zc}
\eeq
where
\beq
     L_{\rm B\leftarrow A}(\l)=\left( \begin{array}{cc}
\psia^{-\half}\psib^{\half}&\l\psia^{-\half}\psib^{-\half}\\
                        &\\
\l\psia^{\half}\psib^{\half}&\psia^{\half}\psib^{-\half}
   \end{array} \right).
\eeq
It's basically the same zero curvature representation as the one
invented by Hirota \cite{H} but this time it is truly quantum.

Multiplying the $L$-operators along the initial data saw and adding
the additional diagonal factor to take care of the quasiperiodicity
of $\psi$'s (see \cite{FTB}) we get the monodromy matrix
\beq
   T(\l)=L_{2M-\half}(\kappa\l)L_{2M-\frac{3}{2}}(\kappa^{-1}\l)\ldots
    L_{\frac{3}{2}}(\kappa\l)L_{\half}(\kappa^{-1}\l)C_2^{\sigma_3}
\eeq
with commuting traces $t(\l)=\tr T(\l)$
\beq
   [t(\l),t(\mu)]=0.
\eeq
The last statement is not quite obvious because the $L$-operator
does not seem to fit into the standard ultralocal scheme of QISM.
It's somewhat similar to the $L$-operator of the quantum Volterra
model \cite{V} but involves highly nonlocal variables $\psi$. The
ultralocality is however easy to recover by glueing our $L$-operators
in pairs
\beq
   \begin{array}{rcl}
   \L_n(\l)&=&L_{2n-\half}(\kappa\l)L_{2n-\frac{3}{2}}(\kappa^{-1}\l)\\
   &&\\        &=&\l\left( \begin{array}{cc}
   \l^{-1} w_{2n-1}^{\half}+\l w_{2n-1}^{-\half}&
   \psi_{2n-1}^{\half}(\kappa w_{2n-1}^{\half}
   +\kappa^{-1}w_{2n-1}^{-\half})\psi_{2n-1}^{\half}\\&\\
   \psi_{2n-1}^{-\half}(\kappa^{-1} w_{2n-1}^{\half}
   +\kappa w_{2n-1}^{-\half})\psi_{2n-1}^{-\half}&
   \l w_{2n-1}^{\half}+\l^{-1} w_{2n-1}^{-\half}
   \end{array} \right)\\&&\\
  t(\l)&=&\tr(\L_M(\l)\L_{M-1}(\l)\ldots\L_1(\l)C_2^{\sigma_3}).
   \end{array}
\eeq
Looking at the commutation relations for the operators involved
\beq \begin{array}{rcl}
           \psi_{2n-1}w_{2n-1}&=&q^2 w_{2n-1}\psi_{2n-1}\\&&\\
    {[}\psi_{2m-1},w_{2n-1}{]}&=&0\;\;\;{\rm if}\;\;\;m\neq n \\
    {[}\psi_{2m-1},\psi_{2n-1}{]}&=&{[}w_{2m-1},w_{2n-1}{]}=0.
     \end{array}
\eeq
we notice that $\L_n$ is not only ultralocal (i.e. matrix elements of the
operators $\L_n$ with different numbers $n$ commute)
but it also resembles the $L$-operator of the general lattice $XXZ$
model (compare also with the lattice sine-Gordon $L$-operator \cite{IK})
and indeed one can easily verify that
\beq
   R({\textstyle \frac{\l}{\mu}})\L_n(\l)\otimes\L_n(\mu)=
   \L_n(\mu)\otimes\L_n(\l)R({\textstyle \frac{\l}{\mu}})
\eeq
where $R(\l)$ is the usual $4\times4$ trigonometric $R$-matrix \cite{FLO}.
This makes the commutativity of $t(\l)$ almost evident while the
commutation relation (\ref{eq:cr}) (or the zero curvature
equation (\ref{eq:zc})) allows to establish easily that $t(\l)$ is
conserved by our dynamics
\beq
    \U^{-1} t(\l) \U =t(\l).
\eeq
Direct inspection shows that $t(\l)$ is a polynomial of $\l^2$ of
degree $M$
\beq
    t(\l)=\sum_{k=0}^M I_k \l^{2k}
\eeq
and thus $M+1$ coefficients $I_k$ constitute the number of commuting
integrals of motion required for the complete integrability.
It remains to establish their functional independence; this can
be done by usual methods.

In classical limit these integrals remain nontrivial and their
existence proves the integrability of our system.

\section{Hirota equation}

The equations considered above do not coincide exactly with the
Hirota equations. The correspondence can be achieved by means of
the following change of variables:
on each second diagonal line SW--NE of the (dual) lattice
$$
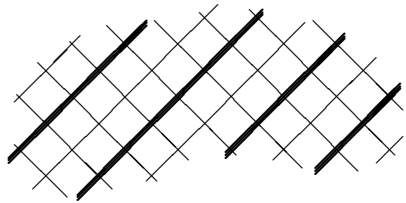
\begin{picture}(160,80)
\thicklines
\put(6,16){\line(1,1){52}}
\put(32,2){\line(1,1){70}}
\put(88,18){\line(1,1){45}}
\put(122,12){\line(1,1){32}}
\put(6,15){\line(1,1){52}}
\put(32,1){\line(1,1){70}}
\put(88,17){\line(1,1){45}}
\put(122,11){\line(1,1){32}}
\put(6,17){\line(1,1){52}}
\put(32,3){\line(1,1){70}}
\put(88,19){\line(1,1){45}}
\put(122,13){\line(1,1){32}}
\thinlines
\put(8,38){\line(1,1){25}}
\put(15,5){\line(1,1){70}}
\put(58,8){\line(1,1){60}}
\put(106,16){\line(1,1){39}}
\put(145,15){\line(1,1){10}}
\put(8,22){\line(1,-1){20}}
\put(9,41){\line(1,-1){36}}
\put(17,53){\line(1,-1){45}}
\put(29,61){\line(1,-1){45}}
\put(43,67){\line(1,-1){49}}
\put(61,69){\line(1,-1){56}}
\put(78,72){\line(1,-1){60}}
\put(97,73){\line(1,-1){55}}
\put(121,69){\line(1,-1){34}}
\end{picture}$$
we are to make change $\psi\rightarrow \psi^{-1}$. If the function
$f(w)$ is ``odd"
\beq
   f(w^{-1})=f^{-1}(w)
\eeq
(the function $f(w)$ from Section 2 is ``odd" while the function
$r(\l,w)$) is ``even": $r(\l,w^{-1})=r(\l,w)$)
then the equations of motion in the new variables (call them $\chi$)
take the form
\beq
    \chin=\chis^{-1} f(q^{-1}\chiw\chie).
\label{eq:chi}
\eeq

Instead of $w$'s we now introduce $u$'s
\beq
   u=\chiw\chie.
\eeq
They are connected exactly in the same way as $\psi$'s and $\chi$'s,
namely, $u$ coincides with either $w$ or $w^{-1}$ on alternating diagonals
of their lattice. The equations of motion in terms of $u$'s read
\beq
  \un=f(q\uw)\us^{-1}f(q^{-1}\ue).
\eeq
Of course on a periodic lattice all these changes are possible only if
the period is even.

The commutation relations for the initial data and the evolution operator
now look as follows:
\beq  \begin{array}{rcl}
    u_{2n\pm 1}u_{2n}& = & q^2 u_{2n}u_{2n\pm 1} \;\;\;\;
                         n=1,2,\ldots,M-1  \\
    u_1 u_{2M}   & = & q^2 u_{2M}u_1       \\
    u_{2M-1} u_{2M}  & = & q^2 u_{2M} u_{2M-1}  \\
    u_m u_n     & = & u_n u_m \;\;\;\;\;\;\;\;\;\;1<|m-n|<2M-1.
     \end{array}
\label{eq:comrel}
\eeq
\beq
    \U=\U_0\prod r(\kappa^2,u_{even})\prod r(\kappa^2,u_{odd})
\eeq
where $\U_0$ is a reflection of all operators $u_n$
\beq
    \U_0^{-1}u_n \U_0=u_n^{-1}.
\eeq
The commutation
relations (\ref{eq:comrel}) are compatible with this inversion
which ensures the existence of $\U_0$.
The conservation laws $I_n$ are even function of $w$'s and so are
commuting with $\U_0$.

It is instructive to remind how equation (\ref{eq:chi}) turns into the
sine-Gordon equation in the continuous limit.
Introducing variables $\varphi$
\beq
   \chi=e^{i\frac{\varphi}{2}}
\eeq
we come to the Hirota form of the equation (\ref{eq:chi})
\beq
   \sin {\textstyle \frac{1}{4}}
   (\varphiw+\varphie-\varphin-\varphis)=
   \kappa^{-2}\sin {\textstyle \frac{1}{4}}
   (\varphiw+\varphie+\varphin+\varphis).
\eeq
Suppose that when the lattice spacing $\Delta$ goes to
zero the variables $\varphi$ turn into the smooth function
$\varphi(t,x)$ and thus
\beq
   \varphiw+\varphie-\varphin-\varphis \longrightarrow
   \frac{\Delta^2}{2}(\partial_x^2\varphi-\partial_t^2\varphi).
\eeq
If now
\beq
   \frac{1}{\Delta^2 \kappa^2}\longrightarrow \frac{m^2}{4}
\eeq
then for $\varphi(t,x)$ we obtain exactly the sine-Gordon equation
\beq
   \partial_t^2\varphi-\partial_x^2\varphi+
   \frac{m^2}{2}\sin 2\varphi=0.
\eeq

Of course, this limiting procedure is valid only in classical case.
The quantum case requires a more delicate balance between the dimensional
parameter $\Delta$ and dimensionless $\kappa^2$, which accounts for
the mass renormalization.

\section*{Conclusion}

Depending on the interests of a reader, one can look upon this paper
from different points of view. For a specialist in dynamical systems
the Hirota model gives a family of examples of symplectic maps of
a finite-dimensional phase space. In particular, as was explicitely
mentioned in \cite{BP} for one degree of freedom we get an integrable
deformation
$$ x_{n+1}-2x_n+x_{n-1}=-2i\ln\frac{\kappa^2+e^{-ix_n}}
                                  {\kappa^2+e^{ix_n}}     $$
of the so-called ``standard map"
$$ x_{n+1}-2x_n+x_{n-1}=-4k\sin x_n.   $$
For related results on symplectic maps see also \cite{MV,PNC}.

For a specialist in quantum field theory the quantized Hirota model
defines a lattice regularization of a relativistic local field-theoretical
model, namely sine-Gordon model. This gives another example of the
relation between alternating chain and relativistic massive models
observed earlier in \cite{FR,DdV}. We give a reasonably explicit
construction of the
evolution operator $\U=e^{-iH\Delta_t}$ where $\Delta_t$
is lattice spacing in time direction and $H$ is the hamiltonian.
Inclusion of $\U$ into algebra generated by local Lax operator allows to
introduce the algebraic Bethe-Ansatz technique for the diagonalization
of $\U$. Here we enter the well developed domain (see \cite{FLO} and
the references therein).

A specialist in exactly solvable plane models of statistical mechanics
should see here new way of writing and arranging of formulas known
in the theory of $Z_n$ model. It is enough to stress that for $q$ being
a root of unity $r(\l,w)$ is exactly the Fateev-Zamolodchikov
$R$-matrix \cite{FZ}.

Finally, a reader interested in the infinite-dimensional algebraic
structures is to see here fragments of a future theory of the algebra
$U_q(\hat{sl}(2))$ \cite{FrR} with which we connect our hopes for the
construction of correlation functions in integrable models of quantum
field theory. This programm is being developed now by the Kyoto
group \cite{M}.

\bigskip {\bf Acknowledgements.} Discussion at different stages of our
work with R. Kashaev, V. Tarasov and M. Semenov-Tian-Shansky were
most stimulating and we are grateful to all of them.

\end{document}